\documentclass[reprint,aps,prx,floatfix,superscriptaddress]{revtex4-2}

\usepackage{hyperref}
\hypersetup{
     colorlinks = true,
     urlcolor   = blue}
\usepackage{xr}
\usepackage{amsmath}
\usepackage{bm}
\usepackage{gensymb}
\usepackage[normalem]{ulem}
\usepackage{euscript}
\usepackage[pdftex]{graphicx}
\usepackage{xspace}
\usepackage{xfrac}
\usepackage{enumerate}
\usepackage{braket}
\usepackage{float,fancyvrb}
\usepackage[T1]{fontenc}
\usepackage{lmodern}
\graphicspath{{./Figures/}}
\usepackage{wrapfig}
\usepackage{comment}
\usepackage{url}
\usepackage{amssymb}
\usepackage{xcolor} 
\usepackage{lineno} 
\usepackage[shortlabels]{enumitem}

\usepackage{mathrsfs}
\usepackage[scr=zapfc, scrscaled=1.15]{mathalfa}

\newcommand{\uu}[1]{\ensuremath{\, \mathrm{#1}}} 

\let\vec\bm

\makeatletter
\let\start@align@nopar\start@align
\let\start@gather@nopar\start@gather
\let\start@multline@nopar\start@multline
\long\def\start@align{\par\start@align@nopar}
\long\def\start@gather{\par\start@gather@nopar}
\long\def\start@multline{\par\start@multline@nopar}
\makeatother

\begin{document}

\baselineskip16pt

\title{Quantum Science and the Search for Axion Dark Matter}

\author{Alexander~O.~Sushkov}
\affiliation{Department of Physics, Boston University, Boston, MA 02215, USA}
\affiliation{Department of Electrical and Computer Engineering, Boston University, Boston, MA 02215, USA}
\affiliation{Photonics Center, Boston University, Boston, MA 02215, USA}


\begin{abstract}
The dark matter puzzle is one of the most important open problems in modern physics. 
The ultra-light axion is a well-motivated dark matter candidate, conceived to resolve the strong-CP problem of quantum chromodynamics. Numerous precision experiments are searching for the three non-gravitational interactions of axion-like dark matter. Some of the searches are approaching fundamental quantum limits on their sensitivity. This Perspective describes several approaches that use quantum engineering to circumvent these limits. Squeezing and single-photon counting can enhance searches for the axion-photon interaction. Optimization of quantum spin ensemble properties is needed to realize the full potential of spin-based searches for the electric-dipole-moment and the gradient interactions of axion dark matter. 
Several metrological and sensing techniques, developed in the field of quantum information science, are finding natural applications in this area of experimental fundamental physics.
\end{abstract}

\maketitle

\clearpage

\section{Dark matter - the signature of physics beyond the standard model} \label{sec:10}

\noindent
Only six parameters are needed to fit all experimental data gathered on the behavior of our universe at cosmological scales~\cite{Spergel2015}. The resulting $\Lambda$CDM model (Lambda cold dark matter) is rich, yet strange. It assumes that general relativity and the Standard Model of particle physics describe the basic physics of the universe throughout its history. Yet, at present time, only $\approx 5\%$ of the universe's energy density is in the form of nucleons, electrons, photons, or neutrinos -- the particles that are described by the standard model~\cite{PDG2020}. Most of the energy in the universe appears to be in the form of \textit{dark energy} that drives the accelerating expansion of the universe. The best-studied possibility is that this acceleration is due to the Einstein cosmological constant $\Lambda$, equivalent to the vacuum energy of empty space~\cite{Peebles2003,Yoo2012}. Most of the matter in the universe is in the form of \textit{dark matter}, which is non-baryonic and has feeble interactions (or none at all) with the standard model particles, aside from gravity.

The early evidence for dark matter long predates the $\Lambda$CDM model. A detailed account of the long history of the dark matter concept can be found in Ref.~\cite{Bertone2018a}. Today, evidence for dark matter is based on numerous observations, including galaxy rotation curves, baryonic acoustic oscillations, and temperature anisotropies of the cosmic microwave background~\cite{Bertone2010,Aghanim2020}. The fact that none of the standard model particles can explain these observations is an indication that dark matter represents evidence for physics beyond the standard model. Since it has only been observed through its gravitational effects, there is a wide range of dark matter candidates, spanning 90 orders of magnitude in mass~\cite{Bertone2018,PDG2019}.

The weakly interacting massive particle (WIMP) is a well-known candidate that has inspired a large number of ultra-sensitive experiments of increasing complexity and scale~\cite{Goodman1985,Bertone2005}. To date, there has been no unambiguous detection, and fundamental backgrounds (neutrino floor) will soon start to limit the sensitivity of direct WIMP searches~\cite{Bertone2010a,Liu2017,Rajendran2017,PDG2020}. Additionally, recent experiments at the LHC, as well as laboratory-scale experiments sensitive to permanent electric dipole moments (EDMs) of atoms and molecules, have placed stringent constraints on theoretical frameworks, such as supersymmetry, that support WIMP dark matter~\cite{Baron2014,Cairncross2017,Khachatryan2017,ACME2018}. Nevertheless, the WIMP concept can be expanded, and broadly-defined WIMPs remain viable dark matter candidates~\cite{Bertone2018}. 

Axions and axion-like particles are in the class of ultra-light dark matter candidates that has attracted increasing attention in the last decade~\cite{Irastorza2018a}. Quantum Chromodynamics (QCD) axions are considered to have the best theoretical motivation, because they solve the strong-CP problem~\cite{Kim2010a}. This problem is briefly formulated as follows. Strong interactions, described by QCD, are allowed to violate the combined charge-conjugation C and spatial inversion (parity) P discrete symmetries. Such violation is quantified by the $\theta$ parameter that enters the QCD Lagrangian, and would give rise, for example, to a permanent EDM of the neutron~\cite{Chupp2019}. Yet, stringent experimental limits on the neutron EDM and other CP-violating strong interactions, indicate that this violation is a factor $\approx10^{10}$ smaller than expected~\cite{Abel2020,Graham2015a}. This constitutes the strong-CP problem. The Peccei-Quinn mechanism solves this problem, by introducing a new scalar field that changes the QCD Lagrangian so that instead of $\theta$ being a constant, it takes on a dynamical effective value which is driven to zero due to spontaneous symmetry breaking~\cite{Peccei1977,Peccei1977a}. The elementary excitation of this field corresponds to the QCD axion, the pseudo-Goldstone boson of the Peccei-Quinn symmetry~\cite{Weinberg1978,Wilczek1978}.

Since its inception, the axion concept has had a deep impact on theoretical physics, emerging naturally from grand unified theories, models with extra dimensions, and string theory~\cite{Svrcek2006}. A plentitude of light pseudoscalar bosons (\textit{axion-like particles, ALPs}) arise readily in string-theory compactifications, as a result of symmetries broken at large energy scales $f_a$~\cite{Svrcek2006,Conlon2006,Arvanitaki2010}. Generic expectations from string theory predict large values for $f_a$, up to the grand unified ($\sim 10^{16}\uu{GeV}$) and Planck ($\sim 10^{19}\uu{GeV}$) scales~\cite{Banks2003,Conlon2006,Cicoli2012,Ernst2018,Graham2018a}.

There are many other dark matter candidates, including relaxions and other scalar fields~\cite{Graham2015b}, hidden photons~\cite{Nelson2011}, superfluid dark matter~\cite{Berezhiani2015}, Planck-scale dark matter~\cite{Adhikari2022}, sterile neutrinos~\cite{Dodelson1994}, and primordial black holes~\cite{Bird2016,Sasaki2016}. Arguments can be made about the degree of theoretical motivation of each specific candidate, but the lack of unambiguous experimental evidence of non-gravitational dark matter interactions signifies the importance of keeping the dark matter search as broad as possible~\cite{DeMille2017,Bertone2018,Rajendran2022}. 

Dark matter is not the only way to resolve puzzles in astrophysical observations, such as galactic rotation curves. Other potential solutions modify the theory of gravitational interactions and dynamics~\cite{McGaugh2014,Chae2020}. 
However these modified gravity theories suffer from several problems, and most scientists in this field pursue the cold dark matter paradigm~\cite{Pardo2020}.

\section{The role of quantum science in fundamental discovery} \label{sec:20}

\noindent
The idea to try to detect non-gravitational interactions of dark matter in a laboratory is many decades old~\cite{Bertone2018a}. Yet, despite numerous sensitive experimental searches of increasing complexity and scale, there has been no unambiguous detection.
One way to guide our efforts towards a potential discovery is to consider recent technological advances that may allow experiments to probe previously unexplored territory. Quantum information science (QIS) is at the forefront of technological innovation~\cite{Monroe2019}. Originating at the foundations of quantum mechanics, the field of quantum science was stimulated by its convergence with information science, when it was realized that quantum mechanical machines can perform certain information-processing tasks faster than any classical computer. 
It is still unknown if and how truly large-scale quantum machines can be built, beyond the noisy intermediate scale quantum (NISQ) devices~\cite{Preskill2018}.
But the concepts and tools of QIS have already inspired a broad spectrum of novel applications, in the fields ranging from life sciences to materials science~\cite{Kucsko2013,Ma2020}. 
Quantum metrology and quantum sensing are already making a significant impact on fundamental physics~\cite{Degen2017}.

\subsection{What is ``Quantum''?}

\noindent
Do we need to define what is meant by ``quantum'', in order to consider how quantum science can be used to search for dark matter? After all, it is the scientific reach of a particular technology that ultimately determines its promise, rather than an arguably arbitrary attribute of ``quantumness''~\cite{Safronova2021}. 
Is a photomultiplier a quantum sensor? It counts photons, quanta of electromagnetic field, with sensitivity beyond the standard quantum limit of a linear amplifier measuring the electromagnetic field. Is a SQUID (Superconducting QUantum Interference Device) a quantum sensor? It makes use of the Josephson effect - the quantum interference of the superconducting wavefunction, split into two paths, interrupted by Josephson junctions. And it has the word ``quantum'' in the acronym.
Answers to such questions are matters of perspective, but, inevitably, they do guide the scope of this Perspective. However, it is important to keep in mind that the ultimate focus is on the potential sensitivity improvements, the feasibility, and the scientific merit of an approach.

Let us loosely define the ``first quantum revolution'', which gave birth to ``quantum 1.0'' technologies, based on lasers, semiconductor, and superconductor devices. The promise of the ``second quantum revolution'' is the development of ``quantum 2.0'' technologies, with the potential to result in impressive performance leaps. For example, a number of approaches aim to evade the Standard Quantum Limit (SQL) of measurement, by making use of entanglement, squeezing, back-action evasion, or strong correlations. More broadly, QIS has developed many ideas for how to optimize preparation, transmission, control, and measurement of correlated quantum states in systems such as spin ensembles, atomic interferometers and clocks, color centers, and superconducting devices.

\subsection{The pioneering efforts}

\noindent
The search for gravitational waves by the Laser Interferometer Gravitational-Wave Observatory (LIGO) is the pioneering example of how ``quantum 2.0'' technologies can make a direct impact on fundamental physics discoveries. Injection of squeezed light into the interferometer improves the photon shot-noise limit to the sensitivity of the Advanced LIGO detectors~\cite{Caves1981}. At frequencies above $50\uu{Hz}$ the sensitivity improvement is up to $3\uu{dB}$, which corresponds to a $40 - 50\%$ increase in the expected gravitational wave event detection rate~\cite{Tse2019}.
At lower frequencies, sensitivity is degraded by quantum back-action, as radiation pressure induces motion of interferometer mirrors. However, frequency-dependent squeezing can achieve a broadband reduction of quantum noise~\cite{Kimble2001,McCuller2020}. 
These quantum technologies are crucial to the scientific reach of the gravitational wave observing programs~\cite{Abbott2020}. 
Indeed there is an intriguing possibility that merger events observed by terrestrial gravitational wave detectors are due to primordial black holes, which could contribute a significant fraction, if not the entirety, of the dark matter abundance~\cite{Bird2016,Sasaki2016,Jedamzik2021,Bhm2021}.

The pioneering efforts of incorporating quantum technologies into gravitational wave detectors have been a key inspiration for the rapid growth of research activity in the field of quantum sensing and metrology~\cite{Degen2017}.
There are also numerous potential applications for quantum sensors, in fields ranging from chemistry and biology, to medicine and geology~\cite{Childress2014}. 
Even within fundamental physics, there are many diverse ideas for how quantum technology can be applied. One example is the development of high-efficiency pattern recognition algorithms, based on quantum annealing, in order to speed up track reconstruction analysis and jet identification in sub-atomic collider experiments~\cite{Bapst2019,Wei2020,Quiroz2021}.
Another area of opportunity is the application of quantum computational resources to simulate high-energy quantum field theories~\cite{Nachman2021}, perform nuclear structure calculations~\cite{Cervia2021}, and model neutrino-nucleus scattering~\cite{Roggero2020}.

This Perspective focuses on a specific area of intense search for new fundamental physics: the use of quantum technologies to accelerate direct searches for ultra-light axion-like dark matter. Many of the ongoing and proposed experiments in this field are approaching, or have already reached, sensitivity levels where quantum resources are needed to achieve their scientific goals. This is also where several of the quantum approaches, having matured in QIS, can find a natural application. Thus there is both a need and an opportunity to use quantum science and engineering. Notably, there are other avenues for sensitivity improvements that are also being pursued, such as development of high-magnetic field technologies~\cite{Battesti2018} and high quality factor resonators~\cite{Jeong2021,DiVora2022,Berlin2022,Giaccone2022}.

\section{Non-gravitational interactions of axion-like dark matter} \label{sec:30}

\noindent 
The Big Bang cosmology of the axion depends on whether the spontaneous breaking of the Peccei-Quinn symmetry occurs before, during, or after inflation~\cite{Linde1988,Linde1991,Tegmark2006,Hertzberg2008}.
Axions can be produced in the early universe non-thermally via the misalignment mechanism~\cite{Preskill1983,Abbott1983,Dine1983}, and via thermal production through axion couplings to the Standard Model plasma~\cite{Masso2002,Salvio2014}.
The misalignment mechanism dominates for large values of the Peccei-Quinn symmetry breaking energy scale, generating a coherent oscillating axion field $a=a_0\cos{(2\pi\nu_at)}$. The oscillation frequency corresponds to the axion Compton frequency $\nu_a=m_ac^2/h$, where $m_a$ is the axion mass, and $h$ is the Planck constant. The field amplitude determines the stored energy density. If we assume that this axion field is the primary component of dark matter, then its amplitude on Earth can be calculated from the local galactic dark matter energy density:
$m_a^2 a_0^2 / 2 = \rho_{\text{DM}} \approx 4 \times 10^{-42}~\text{GeV}^4$~\cite{PDG2019, Graham2013}.
Kinetic energy of the axion-like dark matter field introduces small corrections to its frequency spectrum. The standard halo model predicts the spectral shape with linewidth $(v_0^2/c^2)\nu_a\approx 10^{-6}\nu_a$, where $v_0\approx 220\uu{km/s}$ is
the circular rotation speed of the Milky Way galaxy at the Sun's location~\cite{Turner1990,Evans2019,Gramolin2022}.

Symmetry restricts the interactions that axions and ALPs can have with the particles of the standard model (such as electrons, photons and nuclei)~\cite{Graham2013,Irastorza2018a}. There are three possible interactions.
The defining axion interaction that solves the strong-CP problem is the coupling to the gluon field~\cite{Graham2011}. At low energies this leads to the nucleon electric dipole moment (EDM) interaction
\begin{align}
	\mathcal{H}_{EDM} = g_da\vec{E}^*\cdot\vec{\sigma}/\sigma,
	\label{eq:100}
\end{align}
where $g_d$ is the coupling strength, $\sigma$ is nucleon spin, and $\vec{E}^*$ is an effective electric field~\cite{Budker2014}. This corresponds to a parity- and time-reversal-violating nucleon EDM, given by $d=g_da$.
Since the axion field oscillates at the Compton frequency $\nu_a$, this is not a constant, but an oscillating EDM. 
Recalling the strong-CP problem, we note that this corresponds to an oscillating QCD $\theta$ parameter: $\theta(t) = a(t)/f_a$, with $g_d$ inversely proportional to $f_a$~\cite{Pospelov1999b,Graham2013}.
The EDM coupling generates axion mass, and for the QCD axion $m_a \approx \Lambda_{QCD}^2/f_a$, where $\Lambda_{QCD}\approx 200\uu{MeV}$ is the QCD confinement scale~\cite{Baldicchi2007,Graham2013}.

Axions and ALPs can couple to standard model fermions, such as nucleons or electrons, via the gradient interaction with Hamiltonian
\begin{align}
	\mathcal{H}_{gr} = g_{gr}\vec{\nabla}a\cdot\vec{\sigma},
	\label{eq:200}
\end{align}
where $g_{gr}$ is the coupling strength and $\sigma$ is fermion spin.

The final interaction is with electromagnetic fields, commonly written in terms of the Lagrangian
\begin{align}
	\mathcal{L}_{a\gamma\gamma} = g_{a\gamma\gamma}a\vec{E}\cdot\vec{B},
	\label{eq:300}
\end{align}
where $g_{a\gamma\gamma}$ is the coupling strength and $E,\,B$ are electric and magnetic fields.

\begin{figure}[b!]
	\centering
	\includegraphics[width=\columnwidth]{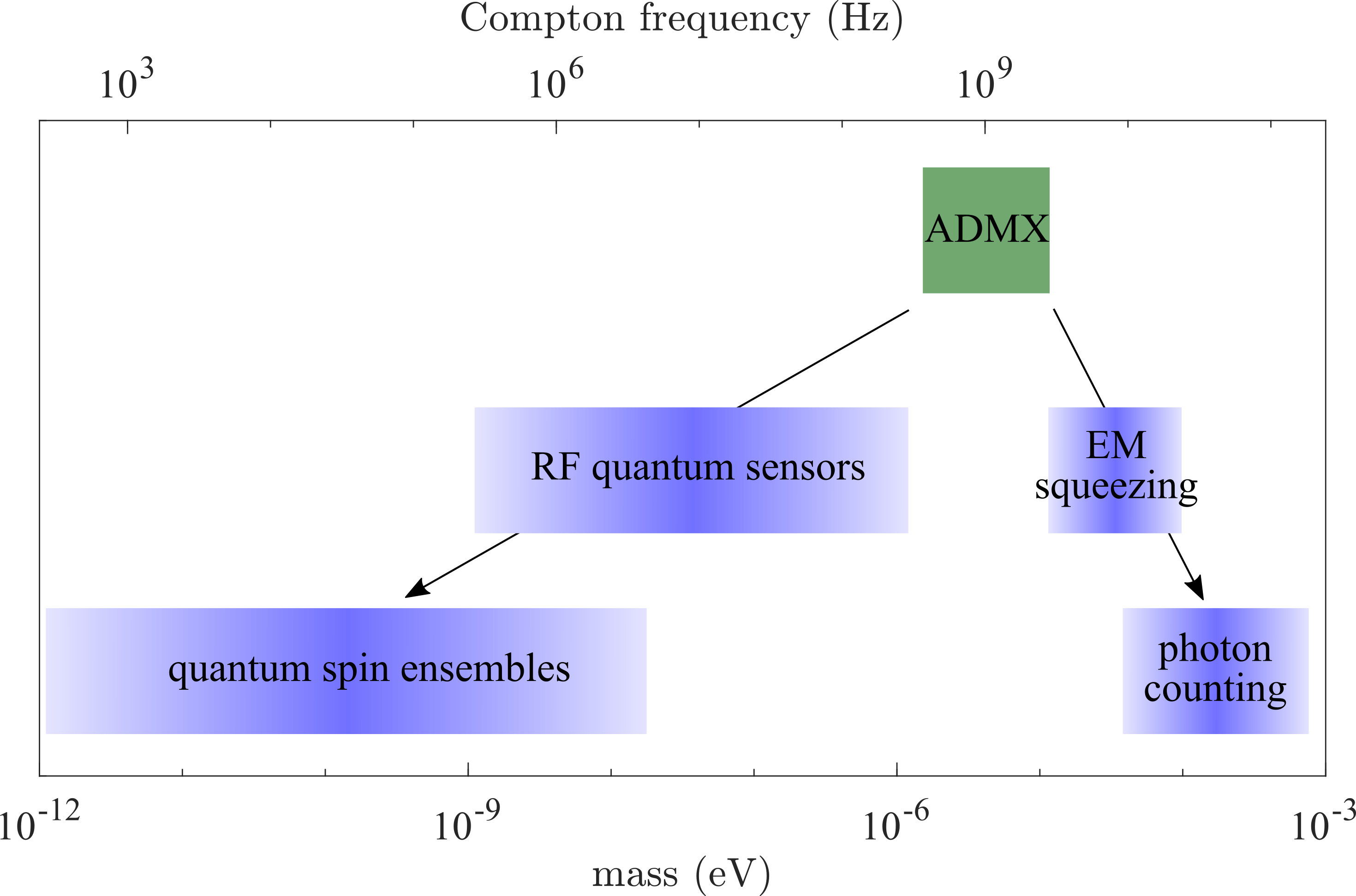}
	\caption{\fontsize{10}{11}\selectfont
		Potential QCD axion dark matter mass range (bottom scale) and corresponding Compton frequencies (top scale). The ADMX experiment aims to cover the mass range between 2.5 - $8.3\uu{\mu eV}$, and future modifications will seek to double the upper end to $16.5\uu{\mu eV}$~\cite{Du2018,Braine2020,Jaeckel2022}. This is only a small fraction of the possible mass range. Approaches based on quantum engineering are an important part of the efforts to broaden the search. Squeezing of quantum fluctuations of the electromagnetic field and photon counting will extend the search at higher masses, and radiofrequency quantum sensors will enable lumped element-based experiments at lower masses. Still lower masses will be explored by experiments that make use of quantum spin ensembles.
	}
	\label{fig:1}
\end{figure}

Generically, axions and ALPs can possess all these couplings.
Direct experimental searches for dark matter can be thought of as transducers, whose goal is to detect the effects of an interaction of a specific dark matter candidate with the experimental apparatus, by converting them into an electromagnetic signal, that is calibrated to give some information about this interaction. 
We consider separately the experiments searching for the electromagnetic interaction of axion-like dark matter with photons, eq.~\eqref{eq:300}, and interaction of axion-like dark matter with spins, eqs.~(\ref{eq:100},\ref{eq:200}).

The possible mass range of axion-like dark matter is extremely broad. An approximate estimate of the mass range where existence of axion-like particles is theoretically and experimentally allowed is $10^{-19}\uu{eV}$ to $10^{-1}\uu{eV}$~\cite{PDG2020}. Focusing on the QCD axion, this window narrows somewhat to $10^{-12}\uu{eV}$ - $10^{-1}\uu{eV}$~\cite{Irastorza2018a}. Only a tiny fraction of this window has been explored in laboratory experiments. The ADMX experiment has searched for the electromagnetic interaction of axion dark matter in the mass range 2.66 - $3.31\uu{\mu eV}$~\cite{Du2018,Braine2020}, and there is a program for extending the upper range to $16.5\uu{\mu eV}$. There are many other ideas for how to cover a much broader swath of axion masses~\cite{Sikivie2021}. Approaches based on quantum engineering are an important part of these efforts, fig.~\ref{fig:1}.

\section{Quantum-enhanced sensors of electromagnetic fields} \label{sec:40}

\begin{figure}[t!]
	\centering
	\includegraphics[width=\columnwidth]{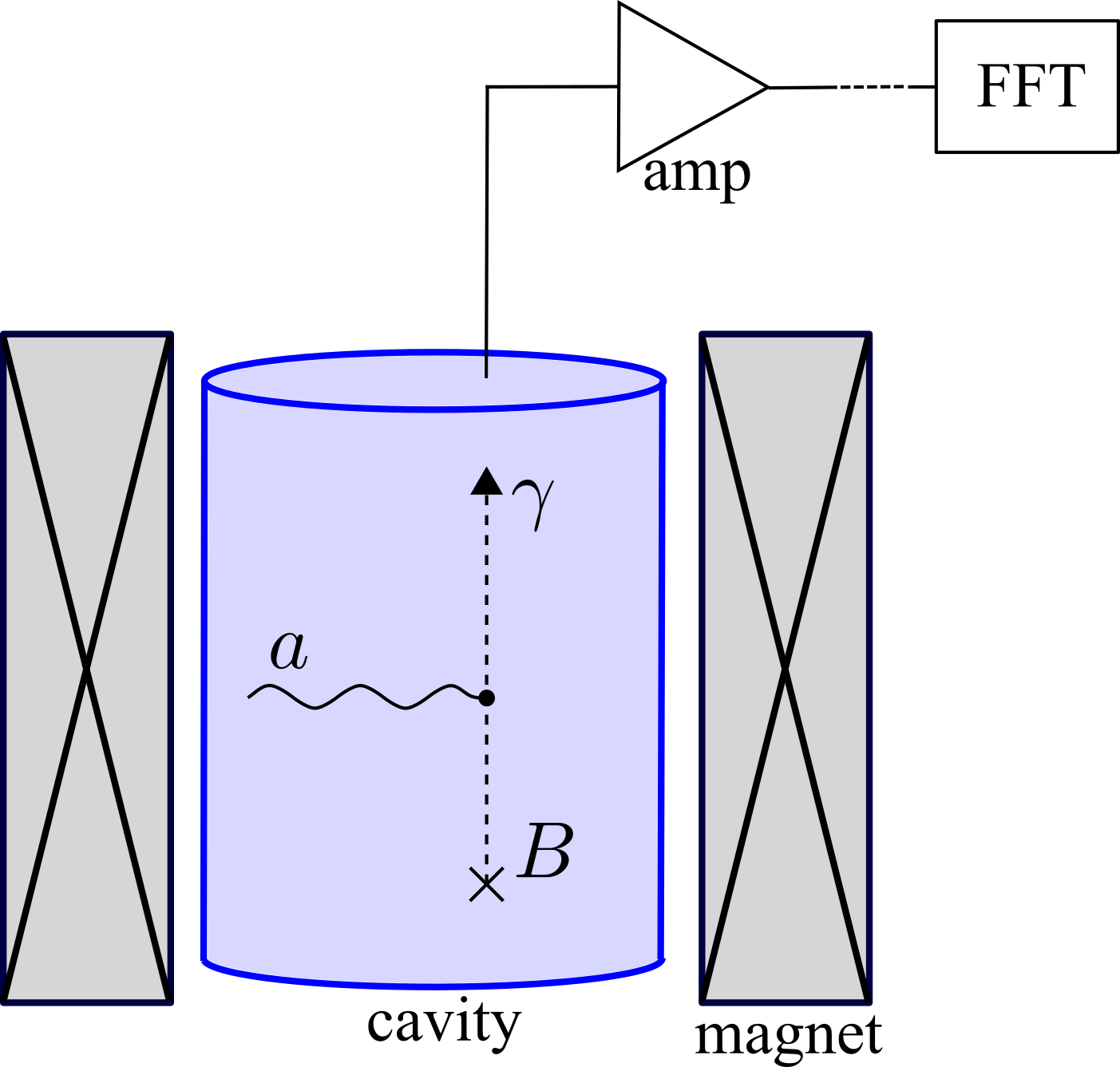}
	\caption{\fontsize{10}{11}\selectfont
		Conceptual schematic of the ADMX haloscope. A resonant cavity is placed in a strong magnetic field created by a superconducting magnet inside a dilution refrigerator. Dark matter axions are converted to photons via the electromagnetic interaction in eq.~\eqref{eq:300}. The resulting electromagnetic signal occupies a narrow band near the axion Compton frequency $\nu_a$. It is coupled out of the cavity and into a sensitive amplifier and detection chain. The search for the unknown axion mass is performed by tuning the cavity resonance.
	}
	\label{fig:ADMX}
\end{figure}

\noindent 
Searches for the electromagnetic interaction of axion-like dark matter are based on the coupling  between the dark matter field and an electromagnetic field sensor, such as a microwave cavity, an optical resonator, or a lumped-element circuit coupled to a sensitive detector~\cite{Sikivie1983}. 
This coupling takes place via the interaction in eq.~\eqref{eq:300}. Microwave cavity-based haloscopes search for conversion of the axion-like dark matter field into photons within a high quality-factor resonant cavity, placed inside a strong magnetic field, fig.~\ref{fig:ADMX}.  
The ADMX experiment is the most mature search of this type. It has achieved a level of sensitivity sufficient to search for dark matter axions with masses between 2.66 and $3.31\uu{\mu eV}$ and,
under the assumption that all of dark matter is in the QCD axion field,
excluded the QCD axion-photon couplings predicted by plausible models for this mass range~\cite{Du2018,Braine2020}. 
A number of cavity-based axion dark matter searches are in development or already exploring ALP masses up to $\approx 50~\mu\text{eV}$~\cite{Graham2015a, Brubaker2017b, McAllister2017, Melcon2018, Alesini2019, Alesini2019a, Lee2020, Backes2021, Kwon2021, Semertzidis2022}. In order to search for lower-mass axions and ALPs coupled to photons, it is possible to use lumped-element circuits~\cite{Sikivie2014b, Chaudhuri2015, Kahn2016, Chaudhuri2018, Ouellet2019, Crisosto2020, Gramolin2021,Salemi2021}. This concept is based on a modification of Maxwell's equations by the $g_{a\gamma\gamma}a\vec{E}\cdot\vec{B}$ interaction: in the presence of a large static magnetic field $B_0$, axion-like dark matter acts as a source of an oscillating magnetic field whose amplitude is proportional to~$B_0$~\cite{Sikivie1983, Wilczek1987}.

\subsection{Electromagnetic squeezing}

\begin{figure*}[t!]
	\centering
	\includegraphics[width=\textwidth]{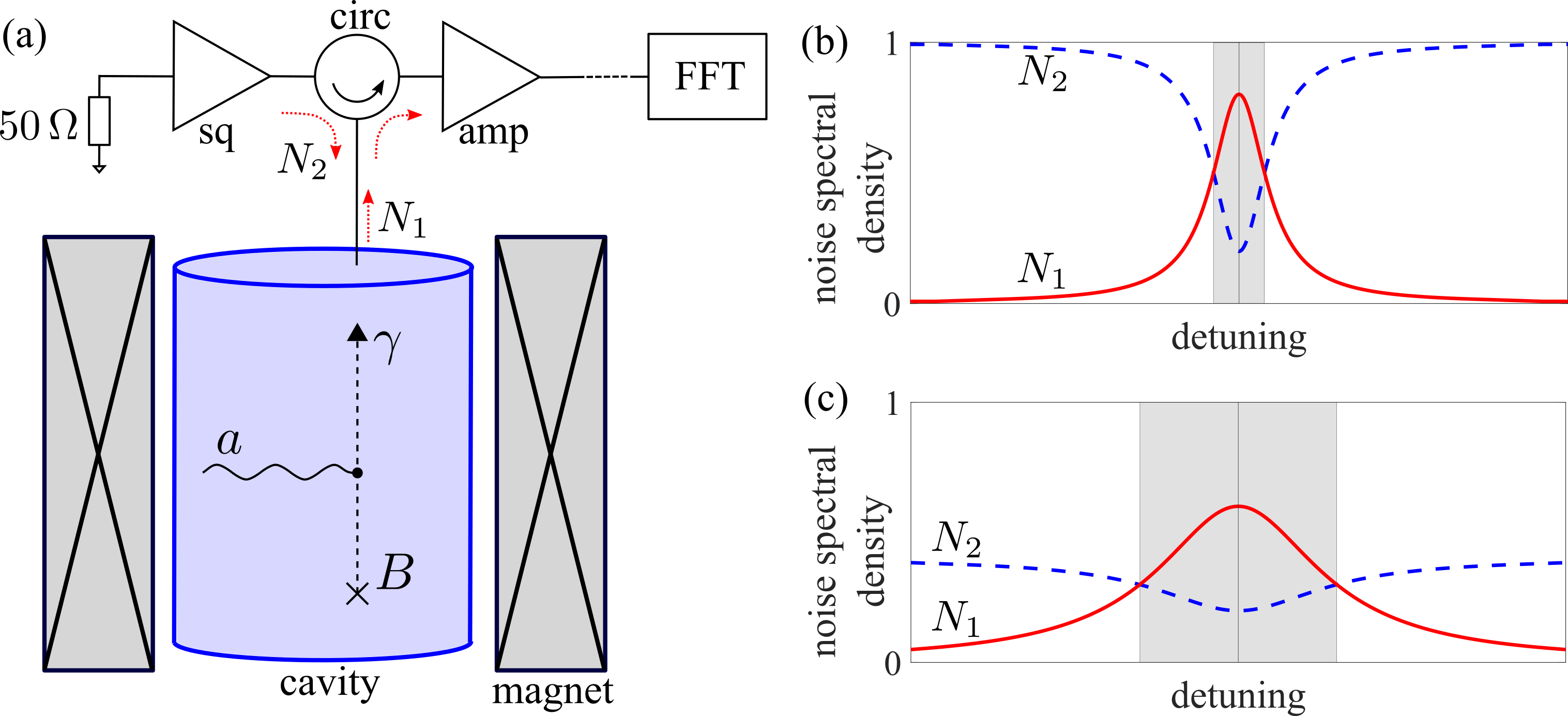}
	\caption{\fontsize{10}{11}\selectfont
		(a) Conceptual schematic of the HAYSTAC haloscope. The detection scheme is similar to the ADMX, consisting of a resonant cavity in a strong magnetic field. HAYSTAC operates at microwave frequencies, where zero-point quantum fluctuations of the electromagnetic field are the dominant noise source. 
		(b) Frequency-dependence of Johnson-Nyquist noise originating in the cavity ($N_1$) and reflected from the cavity ($N_2$), plotted as a function of detuning from cavity resonance. Noise spectral density is in units of single-quadrature vacuum noise $h\nu_c/4$, where $\nu_c$ is the cavity frequency. The grey band indicate the range of frequencies where internal cavity noise dominates.
		(c) By squeezing one of the quadratures of the vacuum state incident on the cavity, HAYSTAC is able to reduce the reflected noise and overcouple the cavity, which broadens the frequency band over which internal cavity noise dominates (grey region; note that the x-axis spans the same scale as in panel (b)). This increases the sensitive bandwidth at each cavity tuning step, and therefore speeds up the axion search.
	}
	\label{fig:HAYSTAC}
\end{figure*}
\noindent
For a microwave cavity haloscope, the estimate for the axion-induced power that has to be detected at the Compton frequency is $\approx 10^{-24}\uu{W}$~\cite{Lamoreaux2013}. Experiments have to be designed to operate at ultra-cold temperatures, to minimize the impact of thermal noise. Zero-point quantum fluctuations of the electromagnetic field are a fundamental source of noise at the level of $h\nu/2$, where $h$ is Planck's constant and $\nu$ is the frequency of the electromagnetic field mode. This noise limits the reach of microwave cavity haloscopes at frequencies above a few GHz. 
Vacuum squeezing can circumvent this limit, and speed up the frequency sweep rate of a resonant search~\cite{Malnou2019}. This has been implemented by the HAYSTAC collaboration, which demonstrated a doubling of their search rate~\cite{Brubaker2017b,Backes2021}. The HAYSTAC approach can be understood using a noise model with two main contributions: $N_1$ and $N_2$. $N_1$ is the noise sourced by the internal cavity loss, as a consequence of the fluctuation-dissipation theorem. Its spectral shape is Lorentzian, with spectral width determined by the cavity quality factor. $N_2$ is the noise incident on, and reflected from, the cavity. This is the Johnson-Nyquist noise, originating from a $50\,\Omega$ termination, held at cryostat base temperature, fig.~\ref{fig:HAYSTAC} (a)~\cite{Clerk2010}.
In the limit where these two contributions dominate over the noise of the subsequent amplifier chain, the width of the band of potential axion masses that are simultaneously probed for a given cavity-tuning step is determined by the range of frequencies where $N_1$ dominates over $N_2$, fig.~\ref{fig:HAYSTAC} (b).
In HAYSTAC the cavity output port is coupled to two Josephson Parametric Amplifiers (JPAs) via a non-reciprocal element (circulator). The noise $N_2$ is coupled into the cavity via one of the circulator ports. The first JPA (labeled `sq') squeezes one of the quadratures of this vacuum state before it enters the cavity, where it can be displaced by the interaction with the axion dark matter field. The second JPA (labeled `amp') unsqueezes the state after it exits the cavity, amplifying the previously-squeezed quadrature. Squeezing does not affect the ratio of the axion-induced signal $S$ to the intrinsic cavity noise $N_1$, which dominates near the cavity resonance. But it does reduce the contribution from the noise $N_2$, which 
makes it beneficial to overcouple the cavity and broaden the frequency band over which $N_1$ dominates over $N_2$, and therefore the band of potential axion masses that are simultaneously probed for a given cavity-tuning step, fig.~\ref{fig:HAYSTAC} (c). This allows larger tuning steps, and speeds up the search at a given axion coupling sensitivity~\cite{Malnou2019}. The HAYSTAC collaboration search for the electromagnetic interaction of axion dark matter placed stringent limits on axions with mass near $17\uu{\mu eV}$~\cite{Backes2021}.

A closely-related method that has been proposed to speed up an axion search scan is to couple the axion-sensitive cavity to an auxiliary resonant circuit via two-mode squeezing and state-swapping interactions~\cite{Wurtz2021}. We should note that both this approach and single-mode squeezing can enhance the scan rate at lower frequencies, where the axion-sensitive cavity has appreciable thermal photon population~\cite{Malnou2019,Chaudhuri2019b}. Thus they are promising experimental directions in this regime. At higher frequencies, however, the technique of photon counting can completely evade the standard quantum limit applicable to linear measurements of electromagnetic fields.

\subsection{Photon counting}

\noindent
Photon counting surpasses the standard quantum limit by measuring only the amplitude and not the phase of the electromagnetic field. For cavity-based experiments searching for the electromagnetic interaction of axion dark matter this means replacing the linear amplifier and detection chain in fig.~\ref{fig:ADMX} with a photon detector. 
The detector dark count rate becomes the key figure of merit, as the Poisson fluctuations in the background count rate limit the experimental sensitivity.
At optical and infra-red frequencies photo-multipliers, avalanche photodiodes, and superconducting nanowire single-photon detectors (SNSPDs) are well-established technologies~\cite{Hadfield2009,Hochberg2019}. 
For example, SNSPDs have recently been used to place limits on dark photon dark matter in the electron-volt mass range~\cite{Chiles2022}.
However such sensors are not suited to detecting lower-energy microwave quanta. 

One possibility for detecting microwave photons is to use highly-excited Rydberg states in alkali atoms~\cite{Tada1999,Tada2006}. This approach has recently seen renewed interest~\cite{Zhu2022a}.
Another approach, spurred by advances in superconducting device technology, has led to the development of single microwave photon detectors in which input photons are coupled to a superconducting qubit, whose state is measured to detect, and even count, the incoming photons~\cite{Besse2018,Kono2018,Dassonneville2020}. 
Such devices have already found applications for magnetic resonance spectroscopy of small ensembles of electron spins~\cite{Albertinale2021}.
Using microwave photon counting for an ultralight dark matter search was explored in recent work, that implemented quantum nondemolition measurements of cavity photons~\cite{Dixit2021}.
A cavity photon shifts the transition frequency of the transmon qubit due to the qubit-cavity interaction. The Ramsey protocol is repeatedly applied to the qubit, performing a quantum non-demolition cavity photon number parity measurement~\cite{Kono2018,Degen2017}. The resulting detection efficiency is $\approx 41\%$, and the false positive probability, proportional to the dark count rate is $\approx 4\times 10^{-4}$~\cite{Dixit2021}. This detector was used in a proof-of-principle search, which set limits on the kinetic mixing parameter of hidden photon dark matter with mass near $25\uu{\mu eV}$~\cite{Dixit2021}.
However, no axion limits were set, because searching for axion dark matter requires the presence of strong magnetic field, which would interfere with detector operation. A possible way forward is to separate the axion-sensitive cavity, placed in a strong magnetic field, from the detector readout cavity. Achieving this while maintaining low dark current rate and high detection efficiency is the formidable technical challenge that remains to be overcome~\cite{Pankratov2022}.

The squeezing and the microwave photon counting are complementary approaches. Squeezing should be deployed at low frequencies (axion masses), where it can accelerate an axion search even in presence of thermal photons, which would be a deleterious background for a photon counting device. At high frequencies photon counting is more effective. It eliminates noise due to vacuum fluctuations, since no real photons are generated by vacuum. Photon counting is also more robust than squeezing against loss, because vacuum fluctuations in lossy components add elecromagnetic field noise, but do not generate real photons. The crossover frequency depends on the technical of each approach, but can be estimated to be $\approx 10\uu{GHz}$~\cite{Lamoreaux2013}.
Both the squeezing the photon counting approaches need to solve the technical challenges that arise from fundamental incompatibility of superconducting quantum devices with high magnetic fields.

\subsection{Radiofrequency quantum sensors and atomic systems}

\noindent
In the lower mass range, at radio-frequencies, thermal noise dominates over zero-point quantum fluctuations. The cross-over is set by the typical dilution refrigerator experiment temperature $\approx 100 \uu{mK}$, which corresponds to $2\uu{GHz}$. Nevertheless, the experimental sensitivity of lumped-element resonant axion dark matter searches benefits from low readout amplifier noise, all the way down to the standard quantum limit~\cite{Chaudhuri2019}. The reason for this can be understood with the argument similar to the description of the HAYSTAC squeezing approach. Lower readout amplifier noise does not change the on-resonance ratio of a potential axion signal to the thermal noise, but it does increase the sensitivity bandwidth over which thermal noise dominates over readout noise, and therefore speeds up the axion search frequency scan~\cite{Chaudhuri2019b}. SQUIDs have been used as readout amplifiers in several lumped-element axion-like dark matter searches~\cite{Ouellet2019,Gramolin2021,Salemi2021}, and they can be optimized to achieve near-quantum-limited performance~\cite{Kinion2011}. The DM Radio collaboration is developing a Radiofrequency Quantum Upconverter (RQU) device, with tunable noise impedance, and compatibility with high-quality-factor operation of the $LC$-resonator that couples to axion dark matter~\cite{Chaudhuri2015,Chaudhuri2019b}. The ultimate goal is not only to achieve the standard quantum limit on amplification, but to develop an approach that goes beyond this limit, making use of backaction evasion to increase the sensitivity bandwidth.

Atomic systems provide an alternative approach to quantum sensing of electromagnetic fields. As mentioned previously, Rydberg atoms are sensitive detectors of microwave electric fields~\cite{Facon2016,Jing2020}.
A number of experiments have explored entanglement and squeezing in order to make optimal use of quantum resources~\cite{Pezze2018}. One of the applications most relevant for dark matter searches is sensing of radiofrequency electric fields, which can be sourced by hidden photons or the axion-photon interaction. A recent experiment has demonstrated $240\uu{nV\,m^{-1}/\sqrt{Hz}}$ sensitivity, using a two-dimensional trapped-ion crystal with $\approx 150$ ions~\cite{Gilmore2021}. The device measures the center-of-mass motion of the trapped ion crystal, probed via the collective electronic spin of the $^9$Be$^+$ ions. Entanglement between these two degrees of freedom allows measurements that evade quantum back-action and thermal noise, achieving electric field sensitivity $\approx 4\uu{dB}$ below the standard quantum limit. 
The demonstrated sensitivity compares favorably to Rydberg atom electrometers.
Another favorable feature of the trapped ion technology for an axion dark matter search is the inherent presence of the strong magnetic field, that is necessary to convert axion dark matter into electromagnetic field. This is in contrast to superconducting detector technologies, which must be operated near zero magnetic field.
The most significant limitation of the trapped ion approach would be the disadvantage of detecting oscillating electric field, compared to magnetic field, by the suppression factor that scales as the ratio of experimental size and the axion Compton wavelength. Nevertheless, if the trapped-ion platform can be scaled up, it has the potential to be competitive with existing limits on the electromagnetic interaction of axion-like dark matter in the neV range.

\section{Spin sensors} \label{sec:50}

\noindent
Historically, experiments sensitive to axion-photon interaction have dominated searches for ultra-light axion-like dark matter. However, as noted in section~\ref{sec:30}, there are two other non-gravitational interactions, which can be used to detect axion dark matter. These interactions couple axions to spin intrinsic angular momentum of electrons, nucleons, or nuclei. Quantum engineering of spin systems has been an active research direction in the field of quantum science, so it is natural to ask how these tools can aid the search for these interactions.

Both of the interactions in eqs.~(\ref{eq:100},\ref{eq:200}) can be written in the following way:
\begin{align}
	\mathcal{H}_{\sigma} = \hbar\gamma \vec{B}^*\cdot\vec{\sigma},
	\label{eq:400}
\end{align}
where $\gamma$ is the gyromagnetic ratio and $\vec{B}^*$ is an oscillating pseudo-magnetic field that exerts a torque on spin $\vec{\sigma}$. Note that this pseudo-magnetic field does not obey Maxwell's equations; it is sourced by the axion dark matter field. In the case of the gradient interaction $\vec{B}^*_{gr} = g_{gr}\vec{\nabla}a/(\hbar\gamma)$, and in the case of the EDM interaction $\vec{B}^*_{EDM} = g_da\vec{E}^*/(\hbar\gamma \sigma)$~\footnote{There are also ideas for searches where $\vec{B}^*$ arises from the axion electromagnetic interaction~\cite{Arza2021}}. In both cases $\vec{B}^*$ oscillates at the Compton frequency $\nu_a$ of the axion field. The objective of experiments that search for these interactions is to detect the effect of this oscillating pseudo-magnetic field on the evolution of the spin ensemble under study.

There are two approaches: detecting the energy deposited by the dark matter field via spin flips, and detecting coherent evolution or energy shift of spins as a result of the effective interaction in eq.~\eqref{eq:400}. The QUAX experiment takes the former approach, measuring the energy deposited by electron spin flips in permeable material yittrium iron garnet (YIG) inside a microwave cavity~\cite{Barbieri2017,Crescini2018,Crescini2020}. There are related experiments and proposals, making use of interaction with magnons and with atomic electrons in rare earths~\cite{Braggio2017,Chigusa2020,Ikeda2022}. The coherent approach is typically used to search for lower axion masses, in experiments that use nuclear spins. The Cosmic Axion Spin Precession Experiments (CASPEr) use nuclear magnetic resonance to search for the EDM and the gradient interactions of axion-like dark matter with nuclear spin ensembles~\cite{Budker2014,Garcon2018,Wang2018,Wu2019a,Garcon2019b,JacksonKimball2020,Aybas2021a}. The closely-related co-magnetometer experiments search for the gradient interaction with nuclear spins~\cite{Vasilakis2009,Jiang2021,Bloch2021}.
Stringent limits on interaction~\eqref{eq:400} at low axion masses can also be extracted from analysis of static EDM experimental data~\cite{Abel2017a,Roussy2021}.

Most of the existing spin-based experimental approaches can be thought of as extensions of an experimental scheme that searches for the axion electromagnetic coupling, but with a spin ensemble acting as the transducer between the axions and the electromagnetic detector. For instance, the QUAX experimental scheme is similar to a microwave cavity axion search shown in fig.~\ref{fig:ADMX}, with a YIG spin sample placed inside the cavity~\cite{Crescini2020}. The CASPEr experimental scheme is similar to a lumped-element circuit search at radio-frequencies, with a nuclear spin sample acting as the transducer. In co-magnetometer experiments the role of the electromagnetic field detector is often played by a laser polarimeter. On the one hand, using spins as transducers increases experimental complexity, since in addition to all the technical parameters in an electromagnetic search (such as sensor noise and coupling), one needs to calibrate and optimize spin ensemble characteristics. On the other hand, this creates more freedom and opportunity to utilize materials and spin ensemble control techniques that can dramatically improve sensitivity. 

\begin{figure}[t!]
	\centering
	\includegraphics[width=\columnwidth]{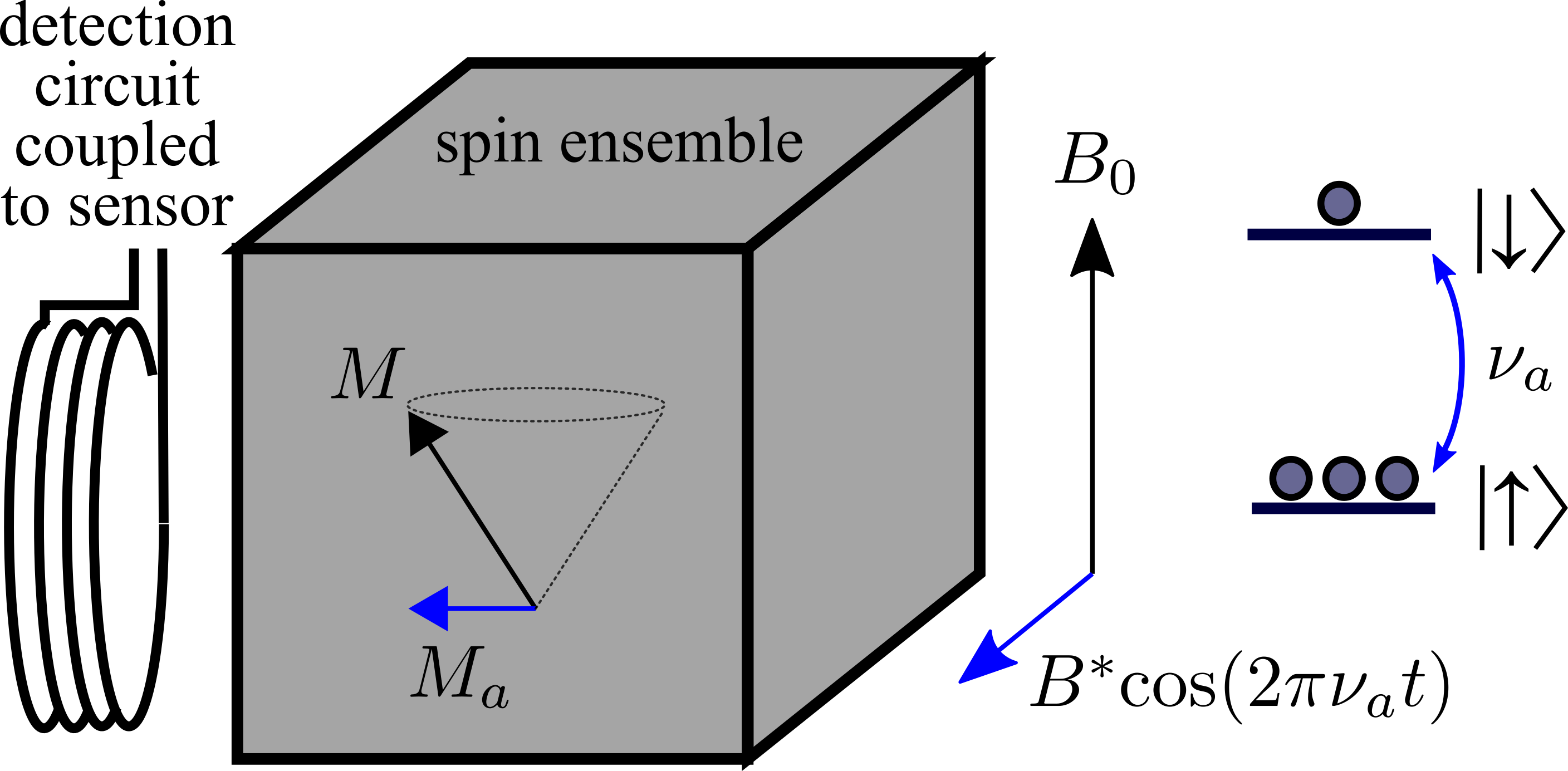}
	\caption{\fontsize{10}{11}\selectfont
		Conceptual schematic of the CASPEr search for the EDM and gradient interactions of axion-like dark matter. The axion-like dark matter field interacts with the spin ensemble via the oscillating effective magnetic field $B^*$. Experiments search for the resulting tilt and precession of the spin ensemble magnetization $M$, when the spin Larmor frequency is on resonance with the axion Compton frequency $\nu_a$. The search for the unknown axion mass is performed by varying the leading field $B_0$, which tunes the spin Larmor frequency.
	}
	\label{fig:CASPEr}
\end{figure}

The interaction Hamiltonian \eqref{eq:400} gives rise to a torque on each spin, whose magnitude is quantified by the Rabi frequency $\Omega_a=\gamma B^*/2$. 
In a resonant coherent detection experiment, the spin ensemble is placed in an external bias magnetic field $B_0$. Resonance occurs if the spin Larmor frequency $\gamma B_0/(2\pi)$ matches the oscillation frequency of the pseudo-magnetic field $B^*$, which is the axion Compton frequency $\nu_a$. 
If the initial spin orientation is along the bias magnetic field $B_0$, then the torque tilts them away from this direction. 
The experimental observable is the oscillating transverse magnetization
\begin{align}
	M_a = uM_0\Omega_aT_2\cos{(2\pi\nu_a t)},
	\label{eq:410}
\end{align}
where $M_0=p\hbar\gamma n$ is the equilibrium magnetization of the spin ensemble with polarization fraction $p$ and number density $n$, $T_2$ is its spin coherence time, and $u$ is a dimensionless spectral factor that takes into account the inhomogeneous broadening of the spin ensemble and the detuning between the axion Compton frequency and the spin Larmor frequency~\cite{Aybas2021a}.
One of the ways to measure the transverse magnetization $M_a$, is with a pickup coil that inductively couples the spin ensemble to an electromagnetic sensor, fig.~\ref{fig:CASPEr}. There are other, more complex detection methods, that make use of atomic magnetometers or co-magnetometers~\cite{Garcon2019b,Wu2019a,Bloch2022c}. 

Let us consider the quantum limits on the sensitivity of the scheme described above. There is still the standard quantum limit on the sensitivity of the electromagnetic field sensor that measures the spin magnetization $M_a$. This SQL, and the approaches that can circumvent it, has been discussed in section~\ref{sec:40} in the context of searches for the axion-photon coupling. In addition, there is now another standard quantum limit -- in the context of spins this is called \textit{spin-projection noise} (SPN)~\cite{Bloch1946}.
A simple way to understand SPN is to consider the following thought experiment. Suppose a single spin-1/2 is prepared in the ``spin-up'' state, namely the quantum state with the $s_z=1/2$ spin projection along the z axis. Then a measurement of the $s_y$ spin component is performed. There are two possible outcomes: $s_y=+1/2,-1/2$, and they are equally likely. If this sequence is repeated $N$ times, or the experiment is performed on $N$ uncorrelated spins, then the (random) mean value of the $s_y$ spin component is normally distributed, with mean 0 and standard deviation $\sqrt{N}/2$. The uncertainty in the transverse spin projection corresponds to a $\delta\theta \approx 1/\sqrt{N}$ uncertainty in the polar angle of the spin, or the spin ensemble. In a spin ensemble with number density $n$, there are $N=nV$ spins in volume $V$, and the SPN magnetization is given by $M_{SPN} \approx (\hbar\gamma/V)\sqrt{N} = \hbar\gamma\sqrt{n/V}$.
As a benchmark, for proton nuclear spins in $V=1\uu{cm^3}$ of water, $\mu_0M_{SPN} \approx 10\uu{fT}$.

The existence of SPN has been noted by Bloch in 1946, and it has been measured in macroscopic spin ensembles in a number of nuclear magnetic resonance (NMR) experiments~\cite{Sleator1985,McCoy1989,Gueron1991}. 
For spin-based axion dark matter searches, conversion of SPN limits to the relevant coupling strength, $g_{gr}$ or $g_d$, is analyzed in ref.~\cite{Aybas2021b}. 
Some of the existing experimental efforts are now approaching SPN-limited sensitivity~\cite{Jiang2021,Aybas2021a}. Technical challenges that have to be overcome include the necessity of a low magnetic field noise environment, the requirement of probing a broad range of Larmor frequencies and, when searching for the EDM interaction, the need to work with static solid crystals~\cite{Aybas2021a}. These are unusual regimes from the point of view of NMR spectroscopy, where experiments are usually done at a fixed magnetic field, and solid-state NMR is usually done with magic-angle spinning~\cite{Andrew1958,Polenova2015}. Nevertheless, spin-based axion dark matter searches are likely to reach SPN-limited sensitivity in the next few years~\cite{Aybas2021b}. Accomplishing this will require quantum engineering of the spin ensemble properties, including polarization and coherence time, as well as optimization of the electromagnetic sensors that are used for ensemble evolution readout. For example, the problem of readout back-action on the spin ensemble (somewhat misleadingly called ``radiation damping'' in NMR spectroscopy) will have to be addressed, especially for gradient interaction searches~\cite{Aybas2021b}. Spin-based experiments are likely to benefit from some of the accomplishments described in section~\ref{sec:40}. 
For instance, superconducting microwave photon detectors have already been used for magnetic resonance spectroscopy of small ensembles of electron spins~\cite{Albertinale2021}. 	
In the near future, the RQU devices, with their tunable noise impedance and potential SQL-limited sensitivity, will play an important role in optimized spin ensemble readout at radio-frequencies.

Is it possible to use quantum engineering to go beyond the SPN? This question has inspired intense research, including ideas such as spin squeezing, entangled states, and quantum error correction~\cite{Pezze2018,Frowis2018,Kessler2014}. 
There are also long-standing arguments that assert that many of these approaches can not make significant improvements to the signal-to-noise ratio of an \textit{optimized} experiment with SPN-limited sensitivity~\cite{Auzinsh2004,Chin2012}. However, even if this is true, there may be other ways to engineer spin ensembles in order to improve such experiments. For example, in sec.~\ref{sec:40} we described how squeezing can achieve an improvement in the useful bandwidth of a search for axion-like dark matter, and therefore speed up the search over the axion Compton frequencies.
Specific spin ensemble quantum engineering schemes will be explored in the next 3-5 years. Maximizing sensitivity to small background fields necessitates the use of macroscopic spin ensembles, on the order of one mole, which leads to unprecedented technical challenges (listed in the previous paragraph), compared to much smaller-scale demonstration experiments. Nevertheless, there are grounds for significant optimism, given the flexibility that is created by choosing the spin ensemble with optimal properties, such as coherence, and combining it with state-of-the-art precision sensors and metrology techniques. If fundamental sensitivity limits, imposed by thermal and quantum noise, can be reached, we are likely to see important scientific breakthroughs in spin-based axion dark matter searches within the next 5-10 years.

\section{Outlook}

\noindent
The dark matter puzzle is one of the most compelling leads in our search for physics beyond the standard model. After decades of searching for the WIMP, there is a rapidly growing recognition that experimental efforts should be broadened, and the QCD axion is a very well-motivated target. The nature of experimental searches for axion-like dark matter offers clear opportunities for the powerful toolbox of quantum science to expand the scientific reach of these searches. Some of these opportunities are sketched in fig.~\ref{fig:1}. The ADMX haloscope is searching for the QCD axion, and is projected to cover the 2.5 - $8.3\uu{\mu eV}$ mass range by 2025. Quantum technologies create opportunities to expand the search range by orders of magnitude, potentially $1\uu{peV}$ to $1\uu{meV}$. The approaches covered in this perspective include quantum spin ensembles, whose optimal sensitivity is at lower axion masses, radiofrequency quantum sensors, electromagnetic cavity squeezing and entanglement, and microwave photon counting using qubits. Whether or not one of these searches discovers the axion depends, of course, on its unknown mass and interactions. Perhaps unexpected discoveries will be made. Quantum science is letting us look in places where no one has ever looked before. Whatever the outcome of the search for axion-like dark matter, there is no doubt that development of quantum precision measurement methods will lead to new devices, sensors, and technologies, which may find diverse applications in fundamental and applied science. After all, quantum engineering is already a key driving force behind innovation and scientific progress in disciplines ranging from computer science to systems engineering and the life sciences.

\section*{Acknowledgements}

\noindent
I thank D.~Budker, D.~F.~Jackson Kimball, and S.~K.~Lamoreaux for valuable comments on the manuscript. I acknowledge support by the Simons Foundation grant 641332, the National Science Foundation CAREER grant PHY-2145162, and the U.S. Department of Energy, Office of High Energy Physics program under the QuantISED program, FWP 100667.


%

\end{document}